\documentclass[aps,prb,preprint,floatfix]{revtex4-1}
\usepackage{amsmath,amssymb,epsfig}
\begin{document}

\title{Interaction created effective flat bands in conducting polymers}
% {Exact ground state for four electrons in a two dimensional
% system with hexagonal repeat units}
\author{Zsolt~Gul\'acsi}
%\author{R\'eka~Trencs\'enyi$^{a}$, Konstantin~Glukhov$^{b}$, 
%and Zsolt~Gul\'acsi$^{c}$}
\address{
%$^{(a)}$ Institute for Nuclear Research, Hungarian Academy of Sciences, 
%H-4026 Debrecen, Bem ter 18/c \\
%$^{(b)}$ Institute for Solid State Physics and Chemistry, Uzhgorod National
%University, Voloshyn Street 54, Uzhgorod 88000, Ukraine\\
%$^{(c)}$ 
Department of Theoretical Physics, University of Debrecen, 
H-4010 Debrecen, Hungary}

\date{May 21, 2014}

\begin{abstract}
For a general class of conducting polymers with 
arbitrary large unit cell and different on-site Coulomb repulsion values 
on different type of sites, I demonstrate in exact terms the emergence 
possibility of an upper, interaction created ``effective'' flat band.
This last appears as a consequence of a kinetic energy quench accompanied
by a strong interaction energy decrease, and leads to a non-saturated 
ferromagnetic state. This ordered state clearly differs from the known 
flat-band ferromagnetism. This is because it emerges in a system without 
bare flat bands, requires inhomogeneous on-site Coulomb repulsions values, 
and possesses non-zero lower interaction limits at the emergence of the 
ordered phase.
\end{abstract}
\maketitle

%%%%%%%%%%%%%%%%%%%%%%%%%%%%%%%%%%%%%%%%%%%%%%%%%%%%%%%%%%%%%%%%%%%%%%%%%%%%

\section{Introduction}

%\subsection{Bare bands and effective bands}
The band concept introduced by the band theory represents the foundation of
our understanding of all solid state devices, and has been successfully used to
explain main physical properties of solids. In its original form, the band 
structure theory assumes an infinite and homogeneous system in which the
carriers, without experiencing inter-electronic interactions, but under the
action of a periodic potential, attain their one-particle quantum mechanical
$E_n({\bf k})$ energy. This last defines the bare band structure described by 
the band index $n$ and wave vector ${\bf k}$. The bare bands are in fact 
eigenstates of the non-interacting part of the Hamiltonian $\hat H_0$.
However, the above assumptions are broken in several practical situations, for
example -- neglecting surfaces, interfaces, inhomogeneities --,
when inter-electronic interactions become important, e.g.  in
the case of strongly correlated systems. In such cases, the materials under
consideration, simply cannot be understood in terms of the bare band structure.
This is the reason why, the effects of the inter-electronic interactions, 
as a necessity, 
have been introduced in the calculation of electronic bands in different
ways, especially based on the density functional theory (DFT). DFT tries to 
coopt the electron-electron many-body effects by the introduction of the 
exchange-correlation term in the functional of the electronic density. On its 
turn, the exchange correlation functional can be approximated in different ways,
for example by local density approximation (LDA) \cite{xkohn}, unrestricted 
Hartree-Fock treatment for the localized orbitals (LDA+U)\cite{xanis}, 
Green function techniques by approximating the self energy as a product 
between the Green's function G and a screened interaction contribution W 
(GW approximation)\cite{xgw}, generalized gradient approximation (GGA) 
\cite{xkohn} which goes beyond LDA by taking account of the gradient
corrections to the density, etc. For strongly correlated systems
the DFT methods need to consider the interaction contributions more and more
accurately, hence in this case there are also present special methods, as 
for example those which take into account input 
DFT data in dynamical mean-field treatment (DFT+DMFT) \cite{xdft1,xdftdmft}. 
For such systems is known that tending to exactitude, as for example in EXX
method \cite{xEXX} taking into account exact-exchange, the accuracy of the
deduced result increases.

Besides the approximations  presented above in a non-exclusive enumeration,
there are rare cases when the effects of the interaction on the bare
band structure
can be exactly seen. For example, in integrable case, the Lieb and Wu solution
\cite{xLieb} shows that for 1D itinerant and periodic systems with nearest 
neighbor hoppings and on-site Coulomb repulsion U, for any $U > 0$, Mott gap 
is present in the spectrum in exact terms, i.e. the 
``effective band'' is always gaped in this case. In my knowledge,
such exact results for non-integrable cases presently are not known.
Starting from this observation, based on the aim to provide valuable information
for non-integrable systems in this field, I present in this paper in exact 
terms how U creates an effective flat band in a system with completely 
dispersive bare bands, what is the physical reason of this process, and what
type of consequences emerge. I note that the effect is important because
information related to flat bands are representing 
a real driving force since they appear in a broad class of subjects 
of large interest today, as quantum Hall effect \cite{y0}, spin-quantum Hall
effect \cite{y00}, 
topological phases \cite{y00,y1}, bose condensations \cite{y2}, highly
frustrated systems \cite{Intr22}, delocalization effects \cite{Intr30} or
symmetry broken ordered phases \cite{Intr15}.

However the effect I describe is not exclusively restricted to quasi 1D
systems (see the end of Sect.IV), the demonstration, born from properties
observed in pentagon chain case \cite{Intr25,Intr8}, is presented on 
conducting polymers with arbitrary large unit cell containing a closed 
polygon and side groups. 
Conducting polymers are an important class of organic systems with a
broad application potential at the level of nanodevices in electronics
\cite{Intr1} or medicine \cite{Intr2}. These materials are in fact 
conjugated polymers \cite{Intr3,Intr4}, which, being metallic, 
are intensively analyzed driven by the aim
to produce different known phases emerging in metals at the level of plastic 
materials. The search for plastic ferromagnetism made entirely from nonmagnetic
elements \cite{Intr5,Intr6,Intr7} enrols as well in this intensively studied 
research direction \cite{Intr8}.
Concerning the theoretical interpretations, in the past, in such systems the
inter-electronic interactions were not considered essential \cite{Intr9}.
However, in recent years, it becomes clear that in conducting and
periodic organic systems, the Coulomb interaction between the carriers plays
an important role. For example, the on-site Coulomb 
repulsion may even reach 10 eV \cite{Intr10}, and it was also conjectured that 
in the highly doped region, the Coulomb interaction would be able to stabilize
magnetic order \cite{Intr11}, all this information being considered during
theoretical studies 
\cite{Intr25,Intr8,Intr12,Intr13,Intr14,Intr18,Intr19,Intr20,Intr21,Intr23}.

The high Hubbard repulsion U values always lead to interesting physical
consequences \cite{xLieb,Intr23a,Intr23b,Intr23c}, so the study of these
cases merits special attention. Hence,
in order to consider in the present case properly the inter-electronic 
interaction and to account accurately for the 
correlation effects, we will use here exact methods. Since conducting polymers
are non-integrable systems, the applied technique is special, and will be 
shortly presented below.

The method we use has no connections to Bethe ansatz, and is based on 
positive semidefinite operator properties. The procedure allows the 
non-approximated deduction of the multi-electronic, even particle number 
dependent ground states, and the non-approximated study of the low lying part 
of the excitation spectrum. The method is applicable for quantum mechanical 
interacting many-body systems, being independent on dimensionality and 
integrability. The technique first transforms in exact terms the system 
Hamiltonian ($\hat H$) in a positive semidefinite form $\hat H = \hat P +C_g$,
where $\hat P$ is a positive semidefinite operator, while $C_g$ a scalar. 
For this
step usually block operators ($\hat A_{{\bf i},\sigma}$) are used which represent a
linear combination of canonical Fermi operators acting on the sites of finite 
blocks connected to the lattice site ${\bf i}$, the positive semidefinite 
form being 
preserved by the $\hat A^{\dagger}_{{\bf i},\sigma} \hat A_{{\bf i},\sigma}$ type of 
expressions. I note, that especially in the above system half filling 
concentration region, in treating the Hubbard type of interaction terms, 
positive semidefinite operators of the form $\hat P_{\bf i}=\hat n_{{\bf i},\uparrow}
\hat n_{{\bf i},\downarrow} - (\hat n_{{\bf i},\uparrow}+\hat n_{{\bf i},\downarrow}) - 1$
are also used, which require at least one electron on the site ${\bf i}$ for
their zero minimum eigenvalue.

The transformation $\hat H = \hat P +C_g$ is valid when a specific relationship 
-- called matching equations -- is present connecting the $\hat H$, and 
block operator
parameters (i.e. the numerical prefactors of the linear combination present
in $\hat A_{{\bf i},\sigma}$). The matching equations, representing a coupled 
non-linear complex algebraic system of equations, must be solved first.
The solution provides the expression of block operator coefficients 
and the scalar $C_g$
in function of $\hat H$ parameters, and, the parameter space region 
${\cal{D}}$, where the transformation in positive semidefinite form of
$\hat H$ is valid. After this step, the exact ground state is constructed
by deducing the most general wave vector $|u\rangle$ which satisfies the
equation $\hat P |u\rangle =0$. The procedure merits attention since several 
techniques for solving this last equation are available today \cite{Intr8,
Intr17,Intr24,Intr25,Intr26}. In the third step, the uniqueness of the solution
is demonstrated by concentrating on the kernel $Ker(\hat P)$ of the operator
$\hat P$, [$Ker(\hat P)$ is a Hilbert subspace containing all vectors
$|v\rangle$ with the property $\hat P |v\rangle =0$]. This is done by showing 
that i) $|u\rangle$ is placed inside $Ker(\hat P)$, and ii) all components of 
$Ker(\hat P)$ can be expressed in terms of $|u\rangle$. The uniqueness proof 
works in the degenerate case as well, when  $\hat P |u(m)\rangle =0$ holds,
$m$ is a degeneracy index, and $|u(m_1)\rangle$ and $|u(m_2)\rangle$, in the
case of $m_1 \ne m_2$, are linearly independent. In this situation, for the
uniqueness proof, we must demonstrate that i) $|u(m)\rangle$ is inside of 
$Ker(\hat P)$ for all $m$ values, and ii) all components of $Ker(\hat P)$
can be expressed as linear combinations of $|u(m)\rangle$.

The last step of the method deduces the physical properties of the ground 
state by calculating different relevant and elevated ground state expectation 
values. We note that if the ground state $|\Psi_g\rangle=|u\rangle$, has been 
obtained, the corresponding ground state energy becomes $E_g=C_g$. 
I must underline, that if the ground state and ground state energy can be 
deduced as function of the total number of particles $N$ ($N$ is maintained
constant during the calculation), we can derive as well non-approximated results
relating the low lying part of the excitation spectrum via the particle number 
dependent chemical potential $\mu(N)=E(N)-E(N-1)$. This can be done for example
by deducing the charge gap $g=\delta \mu=\mu(N+1)-\mu(N)$, where $g=0$ 
($g \ne 0$) reflects conducting (insulating) behavior. 

I note that the transformation in positive semidefinite form of the 
Hamiltonian is always possible. This is because Hamiltonians describing 
physical systems have always a spectrum bounded below. If the lower bound of 
the spectrum is denoted by $C_g$, then $\hat H-C_g =\hat P$ is a positive 
semidefinite operator, independent on dimensionality and integrability. 
Since $C_g$ becomes the ground state energy at the end of the calculations, it
is important to note that when the transformation is performed, the explicit 
form of this constant in function of Hamiltonian parameters is not known 
[see Eq.(\ref{Eq5})], hence the technique not requires the {\it a priory}
knowledge of the ground state energy. Furthermore, the starting point of the
method is a fixed Hamiltonian, hence pre-conceptions or starting information 
relating the ground state wave function are not used or needed. This is why,
in the last step of the procedure, the physical properties of
$|\Psi_g\rangle$ must be separately analyzed.  

The above presented technique has allowed to deduce results in 
circumstances unimaginable before in the context of exact solutions as:
periodic Anderson model in one \cite{Orlik},
two \cite{Intr27}, or three \cite{Intr16,Intr17}
dimensions; disordered and interacting systems in two dimensions
\cite{Intr28}; emergence of stripes and droplets in 2D \cite{Intr29}; 
delocalization effect caused by the on-site Coulomb interaction in 2D
\cite{Intr30}; non-Fermi liquid behavior in 3D
\cite{Intr17}; study of non-integrable quadrilateral
\cite{Intr24,Intr26} or pentagon \cite{Intr8,Intr25} chains.

In the present paper, using the technique described above for general chain 
structures, first I rigorously demonstrate the emergence of an 
effective interaction created flat band,
although the system possesses only dispersive bare bands. By analyzing the
consequences, I show that in the studied parameter space region,
ferromagnetism appears in this class of materials. The emergence of 
the ordered phase is entirely driven by a
huge decrease of the interaction energy, which is accompanied by a quench of
the kinetic energy. The kinetic energy quench, is the physical reason
for the appearance of the
effective upper flat band, whose presence is demonstrated 
on quite general grounds in the general case. The effect is 
related to the presence of different $U_{\bf i}$ on-site Coulomb 
repulsion values at different type of sites inside the unit cell. This allows
a redistribution of the double occupancy $d_{\bf i}$ at different type of sites,
such to attain small $d_{\bf i}$ where $U_{\bf i}$ is high, and vice verse
(i.e. by minimizing $\sum_{\bf i} U_{\bf i} d_{\bf i}$ for fixed and given $U_{\bf i}$
and N values), leading to the observed huge interaction energy 
decrease possibility. When $U_{\bf i}$ is homogeneous, the described effect 
disappears. I note that in the presented case, a ferromagnetic state appears 
on the effective flat band, which
differs significantly from the standard flat band ferromagnetism \cite{Intr15}
because: i) emerges in a system without bare flat bands, ii) requires 
inhomogeneous on-site Coulomb repulsion values, and iii) possesses non-zero
lower limits for the interaction when the ordered phase appears. 

The remaining part of the paper is structured as follows: Section II. presents  
the considered chain structures, Section III. transforms the Hamiltonian in 
positive semidefinite form, presents and solves the matching equations in the
most general case. Section IV demonstrates that in the parameter space region
where the transformation of the Hamiltonian in positive semidefinite form is 
possible to be done, an effective (i.e. interaction created) upper flat band
emerges in the spectrum. Section V. describes the ground state
obtained from the positive semidefinite form of the Hamiltonian, while
Section VI. presents the physical properties of the ground state, and 
the physical characteristics present at the emergence of the ordered state. 
Finally, Section VII. containing the summary and conclusions, closes 
the presentation.
  
\section{The chain structures under consideration}

One analyzes below a general polymer chain whose unit cell is schematically
presented in Fig.1. It
contains $m=m_p+m_e$ sites, from which $m_p$ sites are present in a closed 
polygon, $m_e$ sites are external sites connected to the polygon, and one has
$m > 2$. The in-cell numbering of the sites is given by the index 
$n=1,2,...,m$, and the site positions relative to the lattice site 
${\bf i}$ are given by ${\bf r}_n$.

The Hamiltonian of the system can be written as $\hat H = \hat H_0 + \hat H_U$,
where
\begin{eqnarray}
&&\hat H_0 = \sum_{{\bf i},\sigma} [ \sum_{n,n',n>n'} (t_{n,n'} 
\hat c^{\dagger}_{{\bf i}+{\bf r}_n,\sigma} \hat c_{{\bf i}+{\bf r}_{n'},
\sigma} + H.c.) + \sum_{n=1}^m\epsilon_n \hat n_{{\bf i}+{\bf r}_n,\sigma}],
\nonumber\\
&&\hat H_U=\sum_{\bf i} \sum_{n+1}^m U_n \hat n_{{\bf i}+{\bf r}_n,\uparrow}
\hat n_{{\bf i}+{\bf r}_n,\downarrow},
\label{Eq1}
\end{eqnarray}

%%%%%%%%%%%%%%%%%%%%%%%%%%%%%%%%%%%%%%%%%%%%%%%%%%%%%%%%%%%%%%%%%%%%%%%%%%%
% FIG.1
%%%%%%%%%%%%%%%%%%%%%%%%%%%%%%%%%%%%%%%%%%%%%%%%%%%%%%%%%%%%%%%%%%%%%%%%%%%%
\begin{figure} [h]                                                         %
\centerline{\includegraphics[width=8cm,height=6cm]{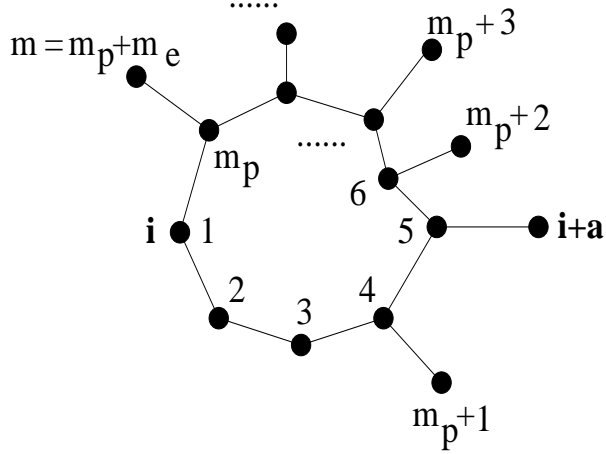}} %
\caption{The unit cell of the chain structures under consideration. 
The cell contains $m=m_p+m_e$ sites, where the $m_p$ sites are included
into a closed polygon, and $m_e$ represents the number of external sites
connected to the polygon. The unit cell is connected to the lattice site
${\bf i}$, while ${\bf a}$ is the Bravais vector. The numbering of sites
inside the unit cell is given by the index $n=1,2,...,m=m_p+m_e$.
Sites $n=1,2,3$ exemplifies sites without external links. The presence of
such sites, or their placement, qualitatively not alters the obtained
results. }      %
\end{figure}                                                               %
%%%%%%%%%%%%%%%%%%%%%%%%%%%%%%%%%%%%%%%%%%%%%%%%%%%%%%%%%%%%%%%%%%%%%%%%%%%%

\hspace*{-0.5cm}where $\hat c^{\dagger}_{{\bf j},\sigma}$ 
creates an electron with spin projection
$\sigma$ at the site ${\bf j}$, $\hat n_{{\bf j},\sigma}=\hat c^{\dagger}_{{\bf j},\sigma}
\hat c_{{\bf j},\sigma}$ represents the particle number operator for the spin
projection $\sigma$ at the site ${\bf j}$, $t_{n,n'}$ are nearest neighbor 
hopping matrix elements connecting the sites ${\bf i}+{\bf r}_{n'}$ and
${\bf i}+{\bf r}_n$, $\epsilon_n$ are on-site one-particle potentials at the
site ${\bf i}+{\bf r}_n$, while $U_n > 0$ are on-site local Coulomb repulsion 
values. I note that during the calculation, periodic boundary conditions are 
used, and $\sum_{\bf i}, \prod_{\bf i},$ (or $\sum_{\bf k}, \prod_{\bf k}$ in momentum
space) mean sums and products, respectively, over $N_c$ cells. The number of 
sites in the system is denoted by $N_{\Lambda}=m N_c$, while the total number of
electrons by $N \leq 2 N_{\Lambda}$. The Hamiltonian parameters are arbitrary, 
but such chosen to not provide bare flat bands (i.e. $\hat H_0$ is without 
non-dispersive bands). Another important remark is related to the $U_n$ Hubbard
interaction values which depend on the particular environment and type of atom
at a given site, hence are different on different type of sites inside the 
unit cell.
Particular cases of the general chain structure from Fig.1 are exemplified
in Fig.2 for pentagon and hexagon cases.
%%%%%%%%%%%%%%%%%%%%%%%%%%%%%%%%%%%%%%%%%%%%%%%%%%%%%%%%%%%%%%%%%%%%%%%%%%%
% FIG.2
%%%%%%%%%%%%%%%%%%%%%%%%%%%%%%%%%%%%%%%%%%%%%%%%%%%%%%%%%%%%%%%%%%%%%%%%%%%%
\begin{figure} [h]                                                         %
\centerline{\includegraphics[width=10cm,height=5cm]{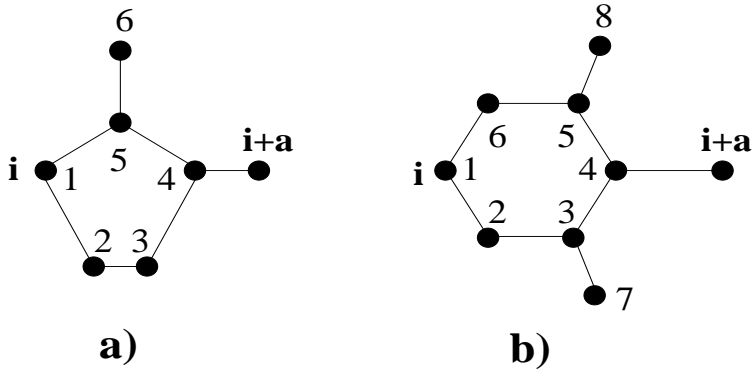}} %
\caption{Exemplifications of the general chain presented in Fig.1 for
a) pentagon ($m_p=5$, $m_e=1$, $m=6$), and b) hexagon ($m_p=6$, $m_e=2$,
$m=8$) cases. The unit cell connected to the lattice site ${\bf i}$ shows 
the $n$ index which provides the in-cell numbering of sites.
${\bf a}$ represents the Bravais vector.}     %
\end{figure}                                                               %
%%%%%%%%%%%%%%%%%%%%%%%%%%%%%%%%%%%%%%%%%%%%%%%%%%%%%%%%%%%%%%%%%%%%%%%%%%%%

Concerning the contributions present in the Hamiltonian from (\ref{Eq1}), 
I note that i) the longer range Coulomb terms have been neglected 
because given by the many-body screening effects, these are much smaller than
the on-site Coulomb repulsion, and ii) the electron-phonon contributions are
not taken into consideration since they become important around half filling,
while the results deduced in this paper are valid far away from half filling, 
in the strongly doped region.

\section{The transformation of the Hamiltonian in positive 
semidefinite form}

\subsection{The transcription of the Hamiltonian}

For the transformation of the Hamiltonian we first introduce at each lattice 
site ${\bf i}$, $m-1=(m_p-2)+(m_e+1)$ blocks on which we 
define block operators as linear combinations of fermionic operators acting
on the sites of the block. The $m-1$ blocks are $m_p-2$ triangles
[namely the triangles constructed on the sites $(m_p,1,2)$, $(m_p,2,3)$,
$(m_p,3,4)$, $(m_p,4,5)$,...,$(m_p,m_{p-1},m_{p-2})$ see Fig.1.],  and
$m_e+1$ bonds [namely, in the case of the unit cell from Fig.1, the
$m_e+1$ bonds connecting the sites $(5,{\bf i}+{\bf a})$, $(4,m_p+1)$, 
$(6,m_p+2)$, $(7,m_p+3)$, ...,$(m_p,m_p+m_e)$]. Consequently, the introduced
$m-1$ block operators become
\begin{eqnarray}
&&\hat G^{\dagger}_{\alpha,{\bf i},\sigma}=a^*_{\alpha,m_p} \hat c^{\dagger}_{{\bf i}+
{\bf r}_{m_p},\sigma}+a^*_{\alpha,\alpha} \hat c^{\dagger}_{{\bf i}+{\bf r}_{\alpha},\sigma} +
a^*_{\alpha,\alpha+1} \hat c^{\dagger}_{{\bf i}+{\bf r}_{\alpha+1},\sigma},
\nonumber\\
&&\hat G^{\dagger}_{m_p-1,{\bf i},\sigma}=a^*_{m_p-1,5} 
\hat c^{\dagger}_{{\bf i}+{\bf r}_{5},\sigma}+a^*_{m_p-1,m+1} 
\hat c^{\dagger}_{{\bf i}+{\bf a},\sigma},
\nonumber\\
&&\hat G^{\dagger}_{m_p,{\bf i},\sigma}=a^*_{m_p,4} 
\hat c^{\dagger}_{{\bf i}+{\bf r}_{4},\sigma}+a^*_{m_p,m_p+1} 
\hat c^{\dagger}_{{\bf i}+{\bf r}_{m_p+1},\sigma},
\nonumber\\
&&\hat G^{\dagger}_{m_{p}+\beta,{\bf i},\sigma}=a^*_{m_p+\beta,\beta+5} 
\hat c^{\dagger}_{{\bf i}+{\bf r}_{\beta+5},\sigma}+a^*_{m_p+\beta,m_p+\beta+1} 
\hat c^{\dagger}_{{\bf i}+{\bf r}_{m_p+\beta+1},\sigma},
\label{Eq2}
\end{eqnarray}
where in the first row $\alpha=1,2,...,m_p-2$  represents the index of the 
block
operators constructed on triangles, the second and third line describes the
first two block operators defined on bonds, namely those constructed on the
site pairs $(4,m_p+1)$, and $(5,{\bf i}+{\bf a})$, and finally, the last row
with the index $\beta=1,2,...,m_e-1$ presents the remaining block operators 
defined on bonds. Furthermore, in Eq.(\ref{Eq2}), the coefficients $a^*_{m,m'}$
are representing the numerical prefactors of the site ${\bf i}+{\bf r}_{m'}$
in the block operator $\hat G^{\dagger}_{m, {\bf i},\sigma}$. I underline 
that if an external bond  
is missing from the polymer, then the corresponding block operator (on the 
respective bond) is also missing from Eq.(\ref{Eq2}). 

Based on the fact that the block operators in Eq.(\ref{Eq2}) are defined in
each unit cell, one constructs the positive semidefinite operator
\begin{eqnarray}
\hat P_{I} = \sum_{{\bf i},\sigma} \sum_{\gamma=1}^{m-1} \hat G_{\gamma,{\bf i},
\sigma} \hat G^{\dagger}_{\gamma,{\bf i},\sigma},
\label{Eq3}
\end{eqnarray}
and introduce the positive semidefinite operator
$\hat P_{\bf j}= \hat n_{{\bf j},\uparrow} \hat n_{{\bf j},\downarrow} -(\hat n_{{\bf j},
\uparrow} + \hat n_{{\bf j},\downarrow}) +1$, (see Sect.I)
which gives rise to the positive semidefinite form
\begin{eqnarray}
\hat P_{II}= \sum_{n=1}^m U_n \hat P_n, \quad \hat P_n = \sum_{\bf i} \hat P_{
{\bf i}+{\bf r}_n}, \quad U_n > 0.
\label{Eq4}
\end{eqnarray}
Using $\hat P_{I}$ and $\hat P_{II}$, the transformed Hamiltonian becomes
\begin{eqnarray}
\hat H = \hat P_{I} +\hat P_{II} + C_g,
\label{Eq5}
\end{eqnarray}
where the scalar $C_g$ has the expression $C_g=q_UN-N_c[\sum_{n=1}^mU_n +
2\sum_{\gamma=1}^{m-1}q_{\gamma}]$. Furthermore, 
$q_{\gamma}=\{ \hat G_{\gamma,{\bf i},\sigma},
 \hat G^{\dagger}_{\gamma,{\bf i},\sigma} \}$, $q_U$ is a scalar which depends
on the parameters of $\hat H$, and can be obtained as a solution of the
matching system of equations Eqs.(\ref{Eq6}-\ref{Eq10}). These
reflect the fact that we transformed 
the starting $\hat H$ from (\ref{Eq1}) dependent on the initial Hamiltonian
parameters $t_{n,n'}, \epsilon_n, U_n$, into $\hat H$ from (\ref{Eq5}) dependent
on block operator parameters $a_{n,n'}$. Consequently, this transformation will 
be valid only if a relationship exists between block operator parameters and the
Hamiltonian parameters. This relationship is fixed by the matching equations
which are obtained as follows: i) one effectuates the calculations in the 
right side of (\ref{Eq5}) obtaining the expression from (\ref{Eq1}), but with 
coefficients dependent on block operator parameters, and ii) taking equal the 
coefficients of the same operator in (\ref{Eq1}) and (\ref{Eq5}).
The results are presented below.

\subsection{The matching equations}

The matching equations preserving the validity of the transformation
of $\hat H$ described above have the 
following structure for the general unit cell presented in Fig.1:\\
\hspace*{1cm}i) For the 
nearest neighbor bonds contained in the polygon and present in the Hamiltonian
via the in-polygon nearest neighbor hopping matrix elements, one obtains
\begin{eqnarray}
-t_{1,m_p}=a^*_{1,1}a_{1,m_p}, \: \: 
-t_{m_p,m_p-1}=a^*_{m_p-2,m_p}a_{m_p-2,m_p-1}, \: \:
-t_{\alpha+1,\alpha}= a^*_{\alpha,\alpha+1}a_{\alpha,\alpha}, 
\label{Eq6}
\end{eqnarray}
where with $\alpha=1,2,...,m_p-2$, one has in total $m_p$ equations.\\
\hspace*{1cm}ii) For the bonds included in the triangular blocks used in the 
construction 
of the block operators, but with zero hopping matrix elements in the 
Hamiltonian from (\ref{Eq1}), one obtains $m_p-3$ equations with
$\alpha'=1,2,...,m_p-3$ :
\begin{eqnarray}
a^*_{\alpha',\alpha'+1} a_{\alpha',m_p} + a^*_{\alpha'+1,\alpha'+1} a_{\alpha'+1,m_p}=
t_{\alpha'+1,m_p}=0.
\label{Eq7}
\end{eqnarray} 
\hspace*{1cm}iii) For the external bonds placed outside of the polygon 
one finds $m_e+1$ equations
\begin{eqnarray}
&&-t_{5,m+1}=-t_{{\bf i}+{\bf r}_5,{\bf i},+{\bf a}}=a^*_{m_p-1,5}a_{m_p-1,m+1}, \: \:
-t_{4,m_p+1}=a^*_{m_p,4}a_{m_p,m_p+1}, 
\nonumber\\
&&-t_{\alpha"+4,m_p+\alpha"}=a^*_{m_p+\alpha"-1,\alpha"+4} a_{m_p+\alpha"-1,m_p+\alpha"},
\label{Eq8}
\end{eqnarray}
where $\alpha"=\beta+1=2,3,...,m_e$.\\
\hspace*{1cm}iv) For the on-site contributions of the sites placed inside 
the polygon one has $m_p$ equations. Namely, by introducing 
$\bar q_U(n)=q_U-(U_n+\epsilon_n)$, one has
\begin{eqnarray}
&&\bar q_U(1)= |a_{1,1}|^2+|a_{m_p-1,m+1}|^2, \quad
\bar q_U(2)= |a_{1,2}|^2+|a_{2,2}|^2, \quad
\bar q_U(3)= |a_{2,3}|^2+|a_{3,3}|^2,
\nonumber\\
&&\bar q_U(4)= |a_{3,4}|^2+|a_{4,4}|^2+|a_{m_p,4}|^2, \quad
\bar q_U(5)= |a_{4,5}|^2+|a_{5,5}|^2+|a_{m_p-1,5}|^2,
\nonumber\\
&&\bar q_U(n)= |a_{n-1,n}|^2+|a_{n,n}|^2+|a_{m_p+n-5,n}|^2, \quad n=6,7,...,m_p-2
\nonumber\\
&&\bar q_U(m_p-1)= |a_{m_p-2,m_p-1}|^2+|a_{m-2,m_p-1}|^2, \quad
\bar q_U(m_p)= \sum_{\alpha=1}^{m_p-2}|a_{\alpha,m_p}|^2+|a_{m-1,m_p}|^2.
\label{Eq9}
\end{eqnarray}
\hspace*{1cm}v) Finally, for the external sites placed outside of the 
polygon and 
representing side groups, with $\gamma=1,2,...,m_e$,
one obtains $m_e$ equations of the form
\begin{eqnarray}
\bar q_U(m_p+\gamma)= |a_{m_p+\gamma-1,m_p+\gamma}|^2,
\label{Eq10}
\end{eqnarray}

The equations Eqs.(\ref{Eq6}-\ref{Eq10}) are representing the matching system of
equations, which contains $M_e=3m_p+2m_e-2$ coupled, non-linear and complex 
algebraic
equations. The unknown variables of this system of equations are the block 
operator coefficients and $q_U$, their number being $M_u=3(m_p-2)+2(m_e+1)+1=
3m_p+2m_e-3$. Since $M_e > M_u$, a supplementary equality remains between the
parameters, which delimits a parameter space region ${\cal{D}}$ where the 
transformation (\ref{Eq5}) is valid.

\subsection{Solution of the matching equations}

The solution technique for the matching equations Eqs.(\ref{Eq6}-\ref{Eq10})
for all $m$ values is similar, and it has practically two steps: 
a) First the equations
connected to hopping matrix elements are used to express unknown parameters 
(i.e. block operator coefficients $a_{n,n'}$) in 
function of other unknown parameters, strongly reducing in this manner the 
number of equations and unknown variables of the problem. b) The expressed
variables are introduced in the remaining equations containing the $U_n$ values.
In the present case, for a) one uses Eqs.(\ref{Eq8},\ref{Eq10}) in order 
to express all coefficients of the $m_e$ block operators 
defined on bonds not touching the site ${\bf i}+{\bf a}$. Using the indices
$\gamma=1,2,...,m_e$ and $\beta=1,2,...,m_e-1$, one finds
\begin{eqnarray}
&&a_{m_p+\gamma-1,m_p+\gamma}= \sqrt{\bar q_U(m_p+\gamma)}, \quad
a_{m_p+\beta,\beta+5}=-\frac{t_{\beta+1,m_p+\beta+1}}{\sqrt{
\bar q_U(m_p+\beta+1)}},
\nonumber\\
&&a_{m_p,4}=-\frac{t_{4,m_p+1}}{\sqrt{\bar q_U(m_p+1)}}, \quad
a^*_{m_p-1,5}=-\frac{t_{5,m+1}}{a_{m_p-1,m+1}},
\label{Eq11}
\end{eqnarray}
where, supplementary, the last equation in the second row is obtained from 
the first equality of (\ref{Eq8}). Based on (\ref{Eq11}), one has $2m_e+1$ 
unknown variables expressed. Now the equations Eqs.(\ref{Eq6},\ref{Eq7}) 
are used in order to provide block operator parameters of the $m_p-2$ block 
operators defined on triangles, as follows: 
For both $\hat G^{\dagger}_{1,{\bf i},\sigma}$ and  $\hat G^{\dagger}_{m_p-2,
{\bf i},\sigma}$ operators, two-two coefficients can be obtained from 
(\ref{Eq6}), namely
\begin{eqnarray}
&&a^*_{1,2}=- \frac{t_{2,1}}{a_{1,1}}, a^*_{1,m_p}=- \frac{t_{1,m_p}}{
a_{1,1}}, 
a^*_{m_p-2,m_p-1}=-\frac{t_{m_p-1,m_p-2}}{a_{m_p-2,m_p-2}}, 
a^*_{m_p-2,m_p}=\frac{t_{m_p,m_p-1}}{t_{m_p-1,m_p-2}} a^*_{m_p-2,m_p-2},
\label{Eq12}
\end{eqnarray}
while for the remaining $m_p-4$ triangles, again from (\ref{Eq6}), one 
coefficient per block operator can be expressed as ($n=2,3,...,m_p-3$).
\begin{eqnarray}
a^*_{n,n+1}=-\frac{t_{n+1,n}}{a_{n,n}}.
\label{Eq13}
\end{eqnarray}
After this step, using (\ref{Eq7}), a second coefficient can be obtained
for the block operators whose prefactors are present in (\ref{Eq13}), namely
($\alpha'=1,2,...,m_p-3$):
\begin{eqnarray}
a_{\alpha'+1,m_p}=-\frac{[\prod_{\alpha=1}^{\alpha'}t_{\alpha+1,\alpha}]t_{1,m_p}}{
a^*_{\alpha'+1,\alpha'+1}
[\prod_{\alpha=1}^{\alpha'}|a_{\alpha,\alpha}|^2]}.
\label{Eq14}
\end{eqnarray}
Now it can be observed that the last (i.e $\alpha'=m_p-3$) equation from 
(\ref{Eq14}), and the last equation from (\ref{Eq12}) express the same 
variable $a_{m_p-2,m_p}$, hence one finds
\begin{eqnarray}
\prod_{\alpha=1}^{m_p-2} |a_{\alpha,\alpha}|^2 =- [\prod_{\alpha=1}^{m_p-2} t_{\alpha+1,
\alpha}] \frac{t_{1,m_p}}{t_{m_p,m_p-1}},
\label{Eq15}
\end{eqnarray}
consequently, for the solution to exists one must has
\begin{eqnarray}
[\prod_{\alpha=1}^{m_p-2} t_{\alpha+1,
\alpha}] \frac{t_{1,m_p}}{t_{m_p,m_p-1}} < 0.
\label{Eq16}
\end{eqnarray}
Equation (\ref{Eq16}) shows that solutions exist only if the product of all 
hopping matrix elements along the closed polygon is a negative number, and 
this result represents one of the conditions which defines ${\cal{D}}$.

In Eqs.(\ref{Eq12}-\ref{Eq15}), 
further $M_2=2m_p-3$ coefficients are expressed, 
while in (\ref{Eq11}) $M_1=2m_e+1$ coefficients are given,
so one has up to this stage $M_1+M_2=2m_p+2m_e-2$ unknown parameters given in 
function of other unknown parameters. Hence one remains with 
$M_u-(M_1+M_2)=m_p-1$ unknown variables
(i.e. $a_{1,1},a_{2,2},...,a_{m_p-3,m_p-3}$, $a_{m_p-1,m+1}$ and $q_U$),
and the remaining $m_p$ matching equations from (\ref{Eq9}). The prefactor
$a_{m_p-2,m_p-2}$ could be expressed in principle from (\ref{Eq15}) (but see
below).

Introducing all the obtained results in (\ref{Eq9}), the remaining $m_p$
matching equations read
\begin{eqnarray}
&&\bar q_U(1)=|a_{1,1}|^2+|a_{m_p-1,m+1}|^2, \: \:
\bar q_U(2)=\frac{t^2_{2,1}}{|a_{1,1}|^2}+|a_{2,2}|^2, \: \:
\bar q_U(3)=\frac{t^2_{3,2}}{|a_{2,2}|^2}+|a_{3,3}|^2,
\nonumber\\
&&\bar q_U(4)=\frac{t^2_{4,3}}{|a_{3,3}|^2}+|a_{4,4}|^2+
\frac{t^2_{4,m_p+1}}{\bar q_U(m_p+1)}, \: \:
\bar q_U(5)=\frac{t^2_{5,4}}{|a_{4,4}|^2}+|a_{5,5}|^2+
\frac{t^2_{5,m+1}}{|a_{m_p-1,m+1}|^2},
\nonumber\\
&&\bar q_U(n)=\frac{t^2_{n,n-1}}{|a_{n-1,n-1}|^2}+|a_{n,n}|^2+
\frac{t^2_{n,m_p+n-4}}{\bar q_U(m_p+n-4)}, \: \: n=6,7,...,m_p-2,
\nonumber\\
&&\bar q_U(m_p-1)=\frac{t^2_{m_p-1,m_p-2}}{|a_{m_p-2,m_p-2}|^2}+
\frac{t^2_{m_p-1,m-1}}{\bar q_U(m-1)},
\nonumber\\
&&q_U-(U_{m_p}+\epsilon_{m_p})=\sum_{\alpha=1}^{m_p-2} |a_{\alpha,m_p}|^2+
\frac{t^2_{m_p,m}}{q_U-(U_{m}+\epsilon_{m})},
\label{Eq17}
\end{eqnarray}
where, in the last line, concerning $|a_{\alpha,m_p}|^2$, for $\alpha=1$ it is 
taken from the second term of (\ref{Eq12}), while for $\alpha \geq 2$ from 
(\ref{Eq14}).

A simple procedure can be applied for solving (\ref{Eq17}). From the second line
from the bottom $|a_{m_p-2,m_p-2}|^2$ can be expressed, then from the third line
from the bottom $|a_{m_p-3,m_p-3}|^2$, similarly, from the fourth line from the 
bottom
$|a_{m_p-4,m_p-4}|^2$, etc. Introducing all these results in
the last line of (\ref{Eq17}) one obtains the equation for $q_U$. 
Eq.(\ref{Eq15}) remains a supplementary condition defining ${\cal{D}}$.
Hence ${\cal{D}}$ will be given by (\ref{Eq15},\ref{Eq16}) and the
conditions $q_U > U_{\alpha}+\epsilon_{\alpha}$, $\alpha=1,2,...,m_p$.

\section{The effective upper flat band created by interaction}

In this section I show that the transformation of the Hamiltonian in 
positive semidefinite form (\ref{Eq5}) together with the solution of the
matching equations presented in subsection III.C. describe in fact 
the emergence of an upper effective (i.e. interaction created) flat
band.

Let us consider that we are placed inside ${\cal{D}}$ in the parameter
space. This means that the matching equations allow solution, consequently
the operators $\hat G^{\dagger}_{\alpha,{\bf i},\sigma}$ exist, are well defined, and 
are $U_n$ dependent as shown by the block operator coefficients expressed for
example in (\ref{Eq11},\ref{Eq17}). In these conditions $\hat P_I$ from
(\ref{Eq3}) entering in the Hamiltonian (\ref{Eq5}) exists and has $U_n$ 
dependence through the $\hat G^{\dagger}_{\alpha,{\bf i},\sigma}$ operators. 
Furthermore, the unique one-particle (i.e. ``kinetic'') contribution in 
(\ref{Eq5}) originates from $\hat P_I$ via 
\begin{eqnarray}
\hat P_{I}= \hat H_{kin} + C_P, \:
\hat H_{kin}= - \sum_{{\bf i},\sigma} \sum_{\alpha=1}^{m-1}
\hat G^{\dagger}_{\alpha,{\bf i},\sigma} \hat G_{\alpha,{\bf i},\sigma}, \:
C_p=2N_c\sum_{\alpha=1}^{m-1} q_{\alpha}, 
\label{Eq18}
\end{eqnarray}
where $C_p$ and $q_{\alpha}$ [defined under (\ref{Eq5})],
are numerical coefficients. 
Consequently the transformed Hamiltonian (\ref{Eq5}) becomes
\begin{eqnarray}
\hat H = \hat H_{kin} + \hat H_{int} + C,
\label{Eq19}
\end{eqnarray}
where for the interaction term one has $\hat H_{int}=\hat P_{II}$, and
$C=C_p+C_g$ is a constant which shifts globally the energy. Since $\hat H_{kin}$
is a one-particle term, it provides a band structure which is ``effective''
because was created by interaction (i.e. depends on the $U_n$ interaction
terms). In what will follows, we will be interested to see what are the
characteristics of the effective band structure created by $\hat H_{kin}$.

Since $m$ sites are present in the unit cell, and the
$\sigma$ index has two values, in ${\bf r}$-space
one has $2mN_c$ different and linearly 
independent $\hat c_{n,{\bf i},\sigma}$, $n=1,2,...,m$ starting  
canonical Fermi operators constructing  $\hat G^{\dagger}_{\alpha,{\bf i},\sigma}$.
The $\hat c_{n,{\bf i},\sigma}$ operators transformed in ${\bf k}$-space provide 
also $2mN_c$ different and linearly independent $\hat c_{n,{\bf k},\sigma}$ 
canonical Fermi operators. Furthermore, one has $m$ bands in the band 
structure, and each band accepts maximum $2N_c$ electrons. 

Now one turns to $\hat H_{kin}$ which contain $2N_c(m-1)$
fermionic operators $\hat G_{\alpha,{\bf i},\sigma}$, $\alpha=1,2,...,m-1$.
Because an $\hat G_{\alpha,{\bf i},\sigma}$ operator, for an arbitrary $\alpha=n_1$,
has at least one site not contained in all $\hat G_{\beta,{\bf i},\sigma}$ operators
with $\beta < n_1$, the block operators $\hat G_{\alpha,{\bf i},\sigma}$
are linearly independent. Transforming them in ${\bf k}$-space, these 
operators lead to $2N_c(m-1)$ different and linearly independent
$\hat G_{\alpha,{\bf k},\sigma}$ operators which however are not canonical
(i.e. $q_{\alpha} \ne 1$, and evidently $q_{\alpha} \ne 0$). Because of this
reason, by normalization to unity, we transform
the $\hat G_{\alpha,{\bf k},\sigma}$ set into a normalized set
obtaining $2N_c(m-1)$ canonical Fermi operators  $\hat C_{\alpha,{\bf k},\sigma}$
(i.e. now, besides
$\{\hat C_{\alpha,{\bf k},\sigma},\hat C_{\alpha',{\bf k}',\sigma'}\}=0,
\{\hat C^{\dagger}_{\alpha,{\bf k},\sigma},\hat C^{\dagger}_{\alpha',{\bf k}',
\sigma'} \} =0$, also $\{ \hat C_{\alpha,{\bf k},\sigma}, 
\hat C^{\dagger}_{\alpha',{\bf k}',\sigma'}\} = \delta_{\alpha,\alpha'} 
\delta_{\sigma,\sigma'} \delta_{{\bf k},{\bf k}'}$ are satisfied). 
Please note that one has $2mN_c$ different starting operators 
$\hat c_{n,{\bf k},\sigma}$,
but the existing $\hat G_{\alpha,{\bf k},\sigma}$ provide only $2N_c(m-1)$
different $\hat C_{\alpha,{\bf k},\sigma}$ operators,
so at this stage $2N_c$ operators $\hat C_{\alpha,{\bf k},\sigma}$ are missing, 
namely those which correspond to $\alpha=m$.

Collecting all the presented information, one has for each 
$\alpha=1,2,...,m-1$  the expression
\begin{eqnarray}
- \sum_{{\bf i},\sigma}
\hat G^{\dagger}_{\alpha,{\bf i},\sigma} \hat G_{\alpha,{\bf i},\sigma}=
\sum_{{\bf k},\sigma}\eta_{\alpha}({\bf k})
\hat C^{\dagger}_{\alpha,{\bf k},\sigma} \hat C_{\alpha,{\bf k},\sigma}.
\label{Eq20}
\end{eqnarray}
Since the left side of (\ref{Eq20}) is negative definite, and 
$\hat C_{\alpha,{\bf k},\sigma}$ are canonical Fermi operators, it results that
$\eta_{\alpha}({\bf k}) < 0$ for all ${\bf k}$ and $\alpha$. 
Indeed, if a state $|\alpha,{\bf k},\sigma\rangle$ state exists with the
property 
$\hat C^{\dagger}_{\alpha',{\bf k}',\sigma'} \hat C_{\alpha',{\bf k}',\sigma'}
|\alpha,{\bf k},\sigma\rangle =\delta_{\alpha',\alpha}\delta_{{\bf k}',{\bf k}}
\delta_{\sigma',\sigma}|\alpha,{\bf k},\sigma\rangle$ and 
$\eta_{\alpha}({\bf k}) \geq 0$,
this would contradict the negative definite nature of the left side of 
(\ref{Eq20}). Consequently, $\hat H_{kin}$ becomes
\begin{eqnarray}
\hat H_{kin}= \sum_{{\bf k},\sigma} \sum_{\alpha=1}^{m-1} \eta_{\alpha}({\bf k})
\hat C^{\dagger}_{\alpha,{\bf k},\sigma} \hat C_{\alpha,{\bf k},\sigma}
\label{Eq21}
\end{eqnarray}
Eq.(\ref{Eq21}) describes $m-1$ effective (interaction dependent) bands
placed at negative energy values, and
$\hat C^{\dagger}_{\alpha,{\bf k},\sigma}$ creates an electron with spin $\sigma$ in 
the $\alpha$th effective band.

Now three steps follow: i) In the knowledge of the $2mN_c$ canonical Fermi 
operators $\hat c_{n,{\bf k},\sigma}$, and $2N_c(m-1)$ canonical Fermi operators
$\hat C_{\alpha,{\bf k},\sigma}$, the remaining $2N_c$ canonical Fermi operators
$\hat C_{\alpha=m,{\bf k},\sigma}$ can be constructed. ii) Since the complete set of
canonical Fermi operators $\hat C_{\alpha,{\bf k},\sigma}$, $\alpha=1,2,...m$,
has been obtained by a linear transformation from the complete set of canonical
Fermi operators $\hat c_{n,{\bf k},\sigma}$, $n=1,2,...,m$, the total particle
number conservation holds and can be written as
\begin{eqnarray}
\sum_{{\bf k},\sigma} \sum_{n=1}^m \hat c^{\dagger}_{n,{\bf k},\sigma}
\hat c_{n,{\bf k},\sigma} = \sum_{{\bf k},\sigma} \sum_{\alpha=1}^m 
\hat C^{\dagger}_{\alpha,{\bf k},\sigma} \hat C_{\alpha,{\bf k},\sigma} = N.
\label{Eq22}
\end{eqnarray} 
iii) Since originates from $\hat H$, $\hat H_{kin}$ from (\ref{Eq21}) must 
describe m bands, but in (\ref{Eq21}) only $m-1$ bands are present. The $m$th
band however can be simply introduced in (\ref{Eq21}) by taking a constant 
$b \geq 0$, multiplying (\ref{Eq22}) by $b$, adding $bN-bN$ to (\ref{Eq21}) and 
introducing the notations 
$\xi_{\alpha<m}({\bf k},b)=\eta_{\alpha<m}({\bf k}) +b, \:
\xi_{\alpha=m}({\bf k},b)=+b=const., \: C_b=-bN$,
based on which, (\ref{Eq21}) becomes
\begin{eqnarray}
\hat H_{kin}= \sum_{{\bf k},\sigma} \sum_{\alpha=1}^{m} \xi_{\alpha}({\bf k},b)
\hat C^{\dagger}_{\alpha,{\bf k},\sigma} \hat C_{\alpha,{\bf k},\sigma} + C_b.
\label{Eq23}
\end{eqnarray}
Eq.(\ref{Eq23}) provides $m$ bands, described by the canonical Fermi 
operators $\hat C^{\dagger}_{\alpha,{\bf k},\sigma}$, while $\xi_{\alpha}({\bf k})$
$\alpha=1,2,...,m$,  is the dispersion relation for the $\alpha$th band, and 
$C_b$ is a ${\bf k}$ independent constant which globally shifts the energies.
As seen from (\ref{Eq23}), given by $\eta_{\alpha}({\bf k})<0$ as shown below
(\ref{Eq20}), $\xi_{\alpha=m}({\bf k})=b=const.$ is the upper band, and it 
is flat.

I further underline that based on (\ref{Eq17}), the emergence of the
effective upper flat band, from mathematical point of view can be interpreted
as a renormalization of the bare $\epsilon_n$ to 
$\epsilon_n^R=\epsilon_n+U_n-q_U$, where $q_U$ is a nonlinear function of all
$U_n$.

As shown above, the interaction created effective upper flat band emerges
on an extremely broad class of polymers. I further note that the effect
exceeds the polymer frame, and appears also in higher dimensions (see for 
example Fig.2 of Ref.[\cite{Intr27}]).

%%%%%%%%%%%%%%%%%%%%%%%%%%%%%%%%%%%%%%%%%%%%%%%%%%%%%%%%%%%%%%%%%%%%%%%%%%%%%

\section{The ground state wave function}

The ground state wave function corresponding to the Hamiltonian presented in
(\ref{Eq5}) for $N=N^*$ number of electrons has the form
\begin{eqnarray}
|\Psi_g(N^*)\rangle = [ \prod_{\sigma} \prod_{\bf i} \prod_{\alpha=1}^{m-1}
\hat G^{\dagger}_{\alpha,{\bf i},\sigma}] \hat F^{\dagger} |0\rangle,
\label{Eq24}
\end{eqnarray}
where $N=N^*=(2m-1)N_c$ (i.e. upper band 
half filled), $|0\rangle$ is the bare vacuum,
and the $\hat F^{\dagger}$ operator introduces one electron with fixed spin 
$\sigma$ in each unit cells in an arbitrary position (i.e.
$\hat F^{\dagger}=\prod_{\bf i} \hat c^{\dagger}_{{\bf i}+{\bf r}_{\bf i},\sigma}$, where
${\bf i}+{\bf r}_{\bf i}$ represents an arbitrary site in the unit cell placed 
at the lattice site ${\bf i}$, and $\sigma$ is fixed). 
The $N=N^*$ expression emerges because one
$\hat G^{\dagger}_{\alpha,{\bf i},\sigma}$ operator introduces one electron in the 
system, one has $2N_c(m-1)$ such operators in (\ref{Eq24}), while 
$\hat F^{\dagger}$ creates $N_c$ electrons, hence 
$N^*=2N_c(m-1)+N_c$. The concentration corresponding to (\ref{Eq24}) is
$n_c=N^*/(2N_{\Lambda})=(2m-1)/(2m)$.

Eq.(\ref{Eq24}) represents the ground state for the following reasons: 
i) $\hat G^{\dagger}_{\alpha,{\bf i},\sigma} \hat G^{\dagger}_{\alpha,{\bf i},\sigma}=0$, 
because the square of an arbitrary linear combination of fermionic operators
is always zero. Consequently, since $\hat G^{\dagger}_{\alpha,{\bf i},\sigma}$
appears both in (\ref{Eq3}) and (\ref{Eq24}), it results that
$\hat P_{I} |\Psi_g\rangle =0$.
ii) Since at fixed $\sigma$,
the operator $[\prod_{\bf i} \prod_{\alpha=1}^{m-1} \hat G^{\dagger}_{\alpha,
{\bf i},\sigma}]\hat F^{\dagger}$ introduces $N_{\Lambda}=mN_c$ electrons in the 
system, one has on each site one electron with spin $\sigma$ present. 
Consequently,
in $|\Psi_g\rangle$, on all sites of the system one has at least one electron, 
and as a consequence $\hat P_{II} |\Psi_g\rangle =0$ is also satisfied
(see the description of the $\hat P_{\bf i}$ in Sect.I.).
In conclusion, $|\Psi_g\rangle$ is the ground state, and the corresponding 
ground state energy is $E_g=C_g$, where $C_g$ is given below (\ref{Eq5}). The 
uniqueness of the solution can be demonstrated on the line of the uniqueness
proof from Ref.\cite{Intr8}. 

I further note that the ground state can be defined also for $N > N^{*}$. In 
this case the ground state expression from (\ref{Eq24}) acquires in its
right side a supplementary product of the form 
$\hat O^{\dagger}=\prod_{\gamma=1}^{N-N^*} c^{\dagger}_{n,{\bf k}_{\gamma},\sigma}$, where
a given, although arbitrary ${\bf k}_{\gamma}$, appears in $\hat O^{\dagger}$ only
once. Since plane wave contributions with fixed spin projection are present 
in $\hat O^{\dagger}$, the ground state at $N > N^*$ becomes a half metallic
conducting state.  

\section{Characteristics of the transition to the ordered state}

The obtained ground state represents a non-saturated ferromagnet.
In order to understand the reasons of the emergence of this state, for fixed 
Hamiltonian parameters placed inside ${\cal{D}}$, one calculates in the
presence of the interaction terms 
different energies (as kinetic energy, interaction energy and total energy) 
using first the non-interacting ground state $|\Psi_{0,g}\rangle$ 
as a trial state, deducing with it $E_{0,kin},E_{0,int}, E_{0,g}=E_{0,kin}+E_{0,int}$.
Then, in a second step, for the same Hamiltonian parameters, one uses
$|\Psi_g\rangle$ from (\ref{Eq24}) in order to deduce the 
exact $E_{kin},E_{int},E_g$. Calculating the relative deviations via
$\delta E_{kin}=(E_{kin}-E_{0,kin})/E_{0,kin}$,
$\delta E_{int}=(E_{int}-E_{0,int})/E_{0,int}$, and 
$\delta E_{g}=(E_{g}-E_{0,g})/E_{0,g}$, expressing these values in percents,
we can analyze
how different energy contributions vary when the ordered state (\ref{Eq24})
emerges. The study has been made on the smallest unit which produces in the 
presented conditions ferromagnetism, namely the two cell system taken with
periodic boundary conditions. 

The obtained results are quite interesting and show that above a given degree
of complexity of the chain situated above a simple triangular chain case
($m=2$, $m_p=2$, $m_e=0$, see Ref.\cite{Intr8}), when the ordered 
state emerges, $\delta E_{kin}$ is 
almost zero (the kinetic energy increases 2-3\%), the interaction energy
strongly decreases (the decrease in $\delta E_{int}$ often reaches almost 70\%),
and as a consequence of these variations, the total energy, described by 
$\delta E_g$, decreases 1-2\%. As it can be seen, the transition to the ordered
phase is clearly driven by the strong decrease of the interaction energy. In 
the same time, the kinetic energy is practically quenched at (or in the close 
vicinity of) $E_{0,kin}$, i.e. the kinetic energy present before the 
interactions have been turned on. These effects appear when differences are
present in the Hubbard interactions at different type of sites inside the unit 
cell, and the behavior disappears when the on-site Coulomb repulsion is
homogeneous, i.e. $U_n=U$ for all $n=1,2,...,m$.    

This behavior can be understood by taking into account that in the studied case
the on-site Coulomb repulsion values are different on different type of sites.
Indeed, in these conditions a supplementary degree of freedom is present for the
decrease of the interaction energy, which is completely missing when the Hubbard
interaction is homogeneous. Namely, the system can reorganize the local double
occupancy $d_n$ such to introduce small $d_n$ where $U_n$ is high and vice 
verse, obtaining a huge interaction energy decrease relative to the interaction
energy values fixed by the double occupancies created by $|\Psi_{0,g}\rangle$.

%%%%%%%%%%%%%%%%%%%%%%%%%%%%%%%%%%%%%%%%%%%%%%%%%%%%%%%%%%%%%%%%%%%%%%%%%%%
% FIG.3
%%%%%%%%%%%%%%%%%%%%%%%%%%%%%%%%%%%%%%%%%%%%%%%%%%%%%%%%%%%%%%%%%%%%%%%%%%%%
\begin{figure} [h]                                                         %
\centerline{\includegraphics[width=9cm,height=6cm]{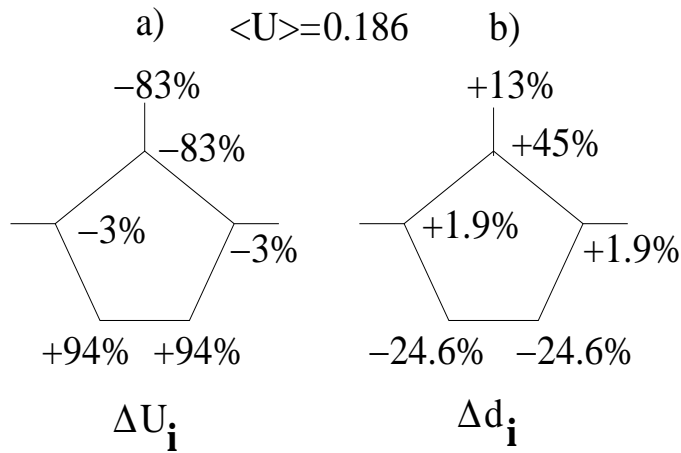}} %
\caption{The modifications $\Delta d_{\bf i}$ in the average local
double occupancy 
$d_{\bf i}= \langle \hat n_{{\bf i},\uparrow} n_{{\bf i},\downarrow}\rangle$ created by 
the interacting ground state $|\Psi_g\rangle$ in (\ref{Eq24}) relative to the
local double occupancy introduced by the non-interacting ground state
$|\Psi_{0,g}\rangle$ (see b)), in function of the local on-site Coulomb 
repulsion 
variations $\Delta U_{\bf i} = U_{\bf i} - \langle U \rangle$ 
on different sites (see
a)), at  $\langle U \rangle=0.186$ in the case of the exemplified chain in
Fig.2.a \cite{ObsFin}.}      %
\end{figure}                                                               %
%%%%%%%%%%%%%%%%%%%%%%%%%%%%%%%%%%%%%%%%%%%%%%%%%%%%%%%%%%%%%%%%%%%%%%%%%%%%

In order to exemplify, I present in Fig.3 results deduced for the $m=6$
case plotted in Fig.2.a inside of its ${\cal{D}}$ region \cite{ObsFin} 
holding in average (in $t=t_{3,4}$ units) $\langle U \rangle = 0.186$
inside the unit cell. Let us consider that one modifies the local
Coulomb repulsion values relative to $\langle U \rangle$ as presented in 
Fig.3.a and calculates the variations in the local average double occupancy
created by (\ref{Eq24}) relative to those double occupancy values, which
were fixed by the non-ordered ground state $|\Psi_{0,g}\rangle$. The results
are presented in Fig.3.b. As can be seen, on the sites where strong U increase
is present (i.e. the local Hubbard interaction is high), the local double 
occupancy in the interacting and ordered 
ground state strongly decreases (the two bottom
sites where in $d_{\bf i}$, 24.6\% decrease is observed). Contrary to this,
on the sites where the Coulomb repulsion values decrease relative to the 
average (hence the local Hubbard interaction is small), 45\% increase
in the double occupancy is observed on the internal site, and a smaller 
increase, but still high (i.e. 13\%, see the top site) is observed in 
$d_{\bf i}$ on the external site. On the sites where $U_{\bf i}$ remains close to 
the average value (in the present case the sites along the line of the chain),
the double occupancy remains almost unchanged (i.e. $\Delta U=-3\%$ produces 
a $\Delta d_{\bf i}=+1.9\%$). I note that given by these modifications 
introduced by
the interacting ground state, the interaction energy decreases almost 
70 \% in the
transition from $|\Psi_{0,g}\rangle$ to $|\Psi_g\rangle$, while the kinetic 
energy increases around 2\% in the same process (i.e. remains practically
unchanged, quenched).

As can be seen, at the emergence of the ordered phase, the system quenches
the kinetic energy, exactly in order to have the possibility to take fully
into account the huge interaction energy decrease possibility, which is 
offered by the non-homogeneous $U$ values inside the unit cell. This kinetic
energy quench is the physical reason of the emergence of an interaction created
effective flat band described in the previous section. 

I underline that when $U_n=U$ is homogeneous, the redistribution of double 
occupancy is no more possible, and the here described mechanism completely
disappears. I further note that the presented mechanism for the emergence of
ferromagnetism is completely different from the Mielke-Tasaki \cite{Intr15}
type of flat band ferromagnetism because in the here described case bare flat 
bands are not present and inhomogeneous $U_n$ values are needed. 
Furthermore, the properties of the ordered phase are
different: for example the conditions leading to ${\cal{D}}$ described at the
end of Sect.III provide lower non-zero limits in $U_n$ for the emergence of
ferromagnetism.

\section{Summary and Conclusions}

The technique used in obtaining the results
allows to deduce particle
number dependent ground states for interacting quantum mechanical many-body 
systems independent on dimensionality and integrability, and also to obtain
non-approximated information relating the low lying part of the excitation 
spectrum. The procedure is based on positive semidefinite operator properties 
and uses successively the following steps: a) transforms the Hamiltonian in a 
positive semidefinite form $\hat H=\hat P + C_g$ where $\hat P$ is a positive 
semidefinite operator and $C_g$ is a scalar, b) deduces the ground state 
$|\Psi_g\rangle$ by constructing the most general solution of the equation
$\hat P |\Psi_g\rangle = 0$. If this equation presents solutions, the 
corresponding ground state energy becomes $E_g=C_g$. c) demonstrates 
the uniqueness of the solution,
and d) analyzes the physical properties of the deduced phase by calculating 
elevated ground state expectation values. The procedure, in principle,
can be applied always, not requiring {\it a priory} information relating the
ground state wave function or ground state energy.

Based on the presented method,
in the high concentration region, a general conducting polymer 
is analyzed, which represents a non-integrable system,
has $m=m_p+m_e > 2$ sites per unit cell, where $m_p$ sites
are included in a closed polygon, and $m_e$ sites, representing side groups, 
are placed outside of it. For the
description a Hubbard type of model is used such that different on-site 
Coulomb repulsion values are allowed at different type of sites inside the 
unit cell, and the system not possesses bare flat bands. In this conditions,
it is rigorously demonstrated that a parameter space region (${\cal{D}}$)
exists where the interactions create an effective upper flat band.

The deduced ground state wave function in ${\cal{D}}$, in the present case,
turns out to be a 
non-saturated ferromagnet. The study of the emergence of the ordered phase 
demonstrates that the transition is entirely driven by a huge decrease of the 
interaction energy, while in the same time, the kinetic energy is quenched. 
The kinetic energy quench  is the physical reason which produces the effective 
flat band. This effect requires a given degree of complexity for the chain,  
and disappears when the Hubbard repulsion becomes homogeneous.
The deduced ferromagnetic state i) appears in the presence of dispersive bare
bands, but demands an interaction created flat band, ii) requires different 
on-site Coulomb repulsion values on different type of sites inside the unit 
cell, iii) leads to non-zero lower limits for the interaction at the emergence 
of the ordered phase, consequently is completely different from the known 
flat-band ferromagnetism.  

\section{Acknowledgments}

The author kindly acknowledges financial support provided by Alexander von 
Humboldt Foundation, OTKA-K-100288 (Hungarian Research Funds for Basic 
Research) and TAMOP 4.2.2/A-11/1/KONV-2012-0036 (co-financed by EU and European 
Social Fund).

%%%%%%%%%%%%%%%%%%%%%%%%%%%%%%%%%%%%%%%%%%%%%%%%%%%%%%%%%%%%%%%%%%%%%%%%%%%
% FIG.1
%%%%%%%%%%%%%%%%%%%%%%%%%%%%%%%%%%%%%%%%%%%%%%%%%%%%%%%%%%%%%%%%%%%%%%%%%%%%
%\begin{figure} [h]                                                         
%\centerline{\includegraphics[width=10cm,height=6.5cm]{2x1G_rendszer_ff.eps}} 
%\caption{a) The studied system. The hexagons, (sites) are labeled by the 
%index $J=I,II,III,IV$, ($n=1,2,...,8$), and sublattices are indicated by 
%black and white dots. b) The torus-like shape of the system taken with
%periodic boundary conditions. c) The used Hamiltonian parameters.}      
%\end{figure}                                                               
%%%%%%%%%%%%%%%%%%%%%%%%%%%%%%%%%%%%%%%%%%%%%%%%%%%%%%%%%%%%%%%%%%%%%%%%%%%%

\bibliographystyle{model1a-num-names}
\bibliography{<your-bib-database>}

\begin{thebibliography}{00}


\bibitem{xkohn}
W. Kohn, Rev. Mod. Phys. {\bf 71}, 1253 (1998).

\bibitem{xanis}
V.I. Anisimov, F. Aryasetianwan and A. I. Lichtenstein, Jour. Phys: Condens.
Matter {\bf 9}, 767 (1997). 

\bibitem{xgw}
F. Aryasetianwan and O. Gunnarsson, Rep. Prog. Phys. {\bf 61}, 237 (1998).

\bibitem{xdft1}
W. Metzner and D. Vollhardt, Phys. Rev. Lett. {\bf 62}, 324 (1989).

\bibitem{xdftdmft} 
A. Georges, G. Kotliar, W. Krauth, and M. J. Rozenberg, 
Rev. Mod. Phys. {\bf 68}, 13 (1996).

\bibitem{xEXX}
M. St\"adele and R. M. Martin, Phys. Rev. Lett. {\bf 84}, 6070 (2000).

\bibitem{xLieb}
E. H. Lieb and F. Y. Wu, Phys. Rev. Lett. {\bf 20}, 1445 (1968).

\bibitem{y0}
Y. Zhang, Y. W. Tan, H. L. Stormer, and P. Kim, Nature {\bf 438}, 201 (2005).

\bibitem{y00}
M. Z. Hasan and C. L. Kane, Rev. Mod. Phys. {\bf 82}, 3045 (2010).

\bibitem{y1}
R. Takahashi and S. Murakami, Phys. Rev. B {\bf 88}, 235303 (2013).

\bibitem{y2}
G. M\"oller and N. R. Cooper, Phys. Rev. Lett. {\bf 108}, 045306 (2012)

\bibitem{Intr22}
O. Derzhko, J. Richter, A. Honecker, and R. Moessner, Phys. Rev. B {\bf 81}, 
014421 (2010).

\bibitem{Intr30}
Z. Gul\'acsi, Phys. Rev. B {\bf 77}, 245113 (2008).

\bibitem{Intr15}
A. Mielke and H. Tasaki, Commun. Math. Phys. {\bf 158}, 341 (1993).

\bibitem{Intr25}
Z. Gul\'acsi, A. Kampf and D. Vollhardt, Phys. Rev. Lett. {\bf 105}, 266403 (2010).

\bibitem{Intr8}
Z. Gul\'acsi, Int. Jour. Mod. Phys. B {\bf 27}, 1330009 (2013).

\bibitem{Intr1}
A. S. Dhoot, {\sl et al.}, Phys. Rev. Lett. {\bf 96}, 246403 (2006).

\bibitem{Intr2}
A. C. R. Grayson, {\sl et al.}, Nature Mater. {\bf 2}, 767 (2003).

\bibitem{Intr3}
R. McNeill, {\sl et al.}, Australian Jour. Chem. {\bf 16}, 1056 (1963).

\bibitem{Intr4}
J. W. van der Horst, P. a. Bobbert and M. A. J. Michels, Phys. Rev. Lett. 
{\bf 83}, 4413 (1999).

\bibitem{Intr5}
O. R. Nascimento, {\sl et al.}, Phys. Rev. B {\bf 67}, 14422 (2003).

\bibitem{Intr6}
F. R. de Paula, {\sl et al.}, Jour. Magn. Magn. Matter. {\bf 320}, 193 (2008).

\bibitem{Intr7}
A. A. Correa, {\sl et al.}, Synth. Met. {\bf 121}, 1836 (2001).

\bibitem{Intr9}
A. J. Heeger et al. Rev. Mod. Phys. {\bf 60}, 781 (1988).

\bibitem{Intr10}
T. O. Wehling, {\sl et al.}, Phys. Rev. Lett. {\bf 106}, 236805 (2011).

\bibitem{Intr11}
G. Brocks, J. Van den Brink and A. F. Morpurgo, Phys. Rev. Lett. {\bf 93},
146405 (2004).

\bibitem{Intr12}
Y. Suwa, {\sl et al.}, Phys. Rev. B {\bf 68}, 174419 (2003).

\bibitem{Intr13}
R. Arita, {\sl et al.}, Phys. Rev. Lett. {\bf 88}, 127202 (2002).

\bibitem{Intr14}
R. Arita, {\sl et al.}, Phys. Rev. B {\bf 68}, 140403(R) (2003).

\bibitem{Intr18}
Z. Gul\'acsi and M. Gulacsi, Phys. Rev. Lett. {\bf 73}, 3239 (1994).

\bibitem{Intr19}
R. Trencs\'enyi, E. Kov\'acs and Z. Gul\'acsi, Phil. Mag. {\bf 89}, 1953 (2009).

\bibitem{Intr20}
R. Trencs\'enyi and Z. Gul\'acsi, Phil. Mag. {\bf 92}, 4657 (2012).

\bibitem{Intr21}
R. Trencs\'enyi, K. Gul\'acsi, E. Kov\'acs, and Z. Gul\'acsi, Ann. Phys. (Berlin)
{\bf 523}, 741 (2011).

\bibitem{Intr23}
R. Trencs\'enyi and Z. Gul\'acsi, Eur. Phys. Jour. B {\bf 75}, 511 (2010).

\bibitem{Intr23a}
M. Gulacsi, H. Van Beijeren and A.C. Levi, Phys. Rev. E {\bf 47}, 2473 (1993);
M. Gulacsi, Phil. Mag. B {\bf 76}, 731 (1997). 

\bibitem{Intr23b}
M. Gulacsi and R. Chan, J. Supercond {\bf 14}, 651 (2001); R. Chan and M. Gulacsi,
Phil. Mag. Lett. {\bf 81}, 673 (2001); R. Chan and M. Gulacsi, Phil. Mag {\bf 84}, 
1265 (2004). 

\bibitem{Intr23c}
D. J. Scalapino, Rev. Mod. Phys. {\bf 84}, 1383 (2012).

\bibitem{Intr16}
Z. Gul\'acsi and D. Vollhardt, Phys. Rev. Lett. {\bf 91}, 186401 (2003). 

\bibitem{Intr17}
Z. Gul\'acsi and D. Vollhardt, Phys. Rev. B {\bf 72}, 075130 (2005).

\bibitem{Intr24}
Z. Gul\'acsi, A. Kampf and D. Vollhardt, Phys. Rev. Lett. {\bf 99}, 026404 (2007).

\bibitem{Intr26}
Z. Gul\'acsi, A. Kampf and D. Vollhardt, Progr. Theor. Phys. Suppl. {\bf 176}, 1
(2008).

\bibitem{Orlik} I. Orlik and Z. Gul\'acsi, Phil. Mag. Lett. {\bf 78}, 177 (1998);
Z. Gul\'acsi and I. Orlik, Jour. of  Phys. A {\bf 34}, L359 (2001).

\bibitem{Intr27}
P. Gurin and Z. Gul\'acsi, Phys. Rev. B {\bf 64}, 045118 (2001);
Z. Gul\'acsi, Eur. Phys. Jour. B {\bf 30}, 295 (2002); 
Z. Gul\'acsi, Phys. Rev. B {\bf 66}, 165109 (2002).

\bibitem{Intr28}
Z. Gul\'acsi, Phys. Rev. B {\bf 69}, 054204 (2004).

\bibitem{Intr29}
Z. Gul\'acsi and M. Gulacsi, Phys. Rev. B {\bf 73}, 014524 (2006).

\bibitem{ut1}
As shown in Ref. \onlinecite{Intr8}, in the simple triangular case, solutions
for $\hat G^{\dagger}_{\alpha,{\bf i},\sigma}$ exist, but not are $U_n$ dependent.
Furthermore, the $|\delta E_{int}|$ and $|\delta E_{kin}|$ variations are
of the same order of magnitude. That is why, a given degree of complexity 
is needed for the chain in order to provide the described mechanism.  

\bibitem{ObsFin} 
One has in the presented case $t=t_{3,4}=t_{4,5}=t_{5,1}=t_{1,2}=1$,
$t_{6,5}=1.2$, $t_{2,3}=-1.1$, $t_{i+a,4}=0.5$, $\epsilon_1=\epsilon_4=-2.5$, 
$\epsilon_2=\epsilon_3=-2.0$, $\epsilon_5=\epsilon_6=-2.1$, and $U_n$ values
expressed in $t$ units.

\end{thebibliography}



\end{document}